%% file: main.tex
\documentclass[prb,twocolumn,longbibliography]{revtex4-2}

\usepackage[utf8]{inputenc}
\usepackage{graphicx}
\usepackage{dcolumn}
\usepackage{bm}
\usepackage{hyperref}
\usepackage{bbold}
\usepackage{amsmath}
\usepackage{amssymb}
\usepackage{commath}
\usepackage{mathtools}
\usepackage{xcolor}
\usepackage{setspace}
\usepackage{comment}
\usepackage[normalem]{ulem}

\DeclareMathOperator{\tr}{tr}

\newcommand{\correlator}{\mathcal{C}}

\newcounter{rev}    
\begin{document}


\title{Fitness landscape for quantum state tomography from neutron scattering}

\author{Tymoteusz Tula}
\author{Jorge Quintanilla}
\email{J.Quintanilla@kent.ac.uk}
\author{Gunnar M\"oller}
\email{G.Moller@kent.ac.uk}
\affiliation{Physics of Quantum and Materials Group, School of Engineering, Mathematics and Physics, University of Kent, Canterbury, Kent CT2 7NH, United Kingdom}

\date{January 24, 2025}

\begin{abstract}
Recently, a direct connection between static structure factors and quantum ground states for two-spin interaction Hamiltonians was proven. This suggests the possibility of quantum state tomography from neutron scattering. Here, we investigate the associated fitness landscape numerically. We find a linear relationship between the mean square distances of structure factors and associated state overlaps, implying a well-behaved fitness landscape. Furthermore, we find evidence suggesting that the approach can be generalized to thermal equilibrium states. We also extend the arguments to the cases of applied magnetic fields and finite clusters.
\end{abstract}

\maketitle

The large number of degrees of freedom in condensed matter systems can make obtaining detailed microscopic insights from experimental data challenging. Usually a model Hamiltonian $\hat{H}$ is used to predict the state of the system in the form of a ground state wave function $\psi_0$ (or, more generally, a thermal density matrix $\rho$). Once the state is known, the expectation value of an observable $\hat{O}$ can be evaluated using a known ground-state functional, $\langle \hat{O}\rangle = \langle \psi_0| \hat{O} | \psi_0 \rangle$ (more generally $\langle \hat{O}\rangle = \mbox{Tr}\rho \hat{O}$). A comparison to the experiment is then used to refine the model $\hat{H}$. In general, however,  the relationships $\hat{H}\to \langle\hat{O}\rangle$ and $\psi_0 \to \langle\hat{O}\rangle$ ($\rho \to \langle\hat{O}\rangle$) are not invertible: there is no guarantee of a unique model $\hat{H}$, or even a unique ground state (or density matrix) compatible with a given observation. A prominent counterexample is provided by Hohenberg and Kohn's theorem, stating that knowledge of the density distribution uniquely determines the ground state of the system\cite{hohenbergInhomogeneousElectronGas1964}, providing the latter is non-degenerate. Generalising this point of view, the prefactors of any unknown terms in the Hamiltonian are fully constrained if the corresponding ground-state or thermal correlation functions are known \cite{ngGeneralizedHohenbergKohnTheorem1991}. Recently, one of us has highlighted that this enables the learning of Hamiltonian parameters of systems with exclusively pairwise interactions of spins where the observable is the ground-state magnetic structure factor $S_{\alpha,\beta}(\mathbf{q})$~\cite{Quintanilla2022}, as well as proving that the many-body wave function can also be inferred directly from the ground-state correlators. The proof of the former, and a weaker form of the latter, have been generalised to finite temperatures~\cite{Murta2022}.

Crucially, the magnetic structure factor is experimentally accessible via diffuse magnetic neutron scattering. This offers the prospect, \emph{in principle}, of using diffuse magnetic scattering on a quantum magnet to completely determine its wave function (quantum state tomography) or its Hamiltonian (Hamiltonian learning). Compared to other schemes aiming to determine states~\cite{Swingle2014, Cramer2010} or Hamiltonians~\cite{Chertkov2018,Greiter2018,Bairey2019,Qi_2019,turkeshiEntanglementGuidedSearchParent2019,Anshu2021} from local measurements that have been discussed in the context of quantum simulators this has the advantage that it does not require the full covariance matrix~\cite{Chertkov2018,Greiter2018,Bairey2019,Qi_2019} and that all the necessary correlators are determined simultaneously in the neutron-scattering measurement so there is no scaling of experimental complexity~\cite{Anshu2021} with the size of the system or the range of interactions. 

The main purpose of the present work is to determine whether the above scheme could work \emph{in practice}. To address this question, we study how the normalised difference between random trial- and a given target scattering function varies as the trial state moves away from the target. The ensemble of r.m.s. differences across a large set of trial states represents a “fitness landscape” for the inverse neutron-scattering problem (according to “Theorem 2” in [3]). We numerically explore both the zero and finite temperature cases. We also generalise previous discussions \cite{Quintanilla2022,Murta2022} to the case where there are applied magnetic fields and provide further evidence of the validity of the theorems at finite temperature.

{\it General framework:}
We first provide key definitions to set up the general framework and choice of models we consider. Specifically, we define appropriate metrics in the spaces of scattering functions and quantum states. We will assume the Hamiltonian is of the class considered in Refs.~\cite{Quintanilla2022,Murta2022}, featuring only two-spin interactions:
\begin{equation} \label{eq:HeisenbergHam}
    H_{J} = \sum_{i,j}\sum_{\alpha, \beta} J_{i,j}^{\alpha, \beta} S_i^{\alpha} S_j^{\beta},
\end{equation}
where operators $S_i^{\alpha}$ represent spin components along axis $\alpha\in{x,y,z}$ at site $i$ of the lattice, and $J_{i,j}^{\alpha, \beta}$ are a completely generic set of coupling constants between these. Note that this form does not include linear terms. Meanwhile, in the experimental setup, one cannot exclude local magnetic fields affecting the state of the system. Indeed such fields are often an important tool to explore the physics of a given material. For instance, they can be used to tune a material to quantum criticality \cite{Coldea2010,Breunig2017}. Our first result is a proof that the class of generic spin Hamiltonians described by Eq.~(\ref{eq:HeisenbergHam}) in fact includes all Hamiltonians with local fields,
\begin{equation} \label{eq:HeisenbergHamField}
    H_{J + h} = \sum_{i,j}\sum_{\alpha, \beta} {J}_{i,j}^{\alpha, \beta} S_i^{\alpha} S_j^{\beta} + \sum_{i}\sum_{\alpha} h_i^\alpha S_i^\alpha,
\end{equation}
if we allow the coupling constants  $J_{i,j}^{\alpha, \beta}$ in Eq.~(\ref{eq:HeisenbergHam}) to be complex. A proof follows straightforwardly by noting that the commutators of products of two spin operators on the same site naturally generate imaginary single-spin terms. We give a detailed proof in the supplementary material \cite{supp-mat}. In our numerical study we will therefore include systems of the form (\ref{eq:HeisenbergHamField}) with real exchange constants  $J_{i,j}^{\alpha, \beta}$  and magnetic fields  $h_i^\alpha$. 

Our methodology of testing the fitness landscape of scattering functions can be summarised as follows: Firstly, we select a reference Hamiltonian $H$ describing a chain of $L$ sites $S=\frac{1}{2}$. Calculating either a ground state $\psi$, or thermal density function $\rho(\beta)$ allows us to evaluate the full diffuse magnetic scattering function $\mathcal{S}(\mathbf{q})$ for the reference Hamiltonian. Then we proceed to add random variations to the ground state/density function: $\psi' = \psi + \delta\psi$, $\rho' = \rho + \delta\rho$ and evaluate the scattering functions $\mathcal{S}'(\mathbf{q})$ that are obtained from the new states. By defining a metric for the distance between two scattering functions $\Delta \mathcal{S}(\mathcal{S}, \mathcal{S}')$ (chosen as a properly normalised mean-square deviation), we can study the relationship of $\Delta \mathcal{S}$ with the distance between the respective states by examining multiple samples of random variations $\delta \psi$ or $\delta \rho$. Overlaps induce a natural metric in state space; for further details see the supplementary material 
\cite{supp-mat}.

{\it State tomography:} 
\begin{figure}
    \centering 
    \includegraphics[width=0.85\columnwidth,trim={0cm 0cm 0cm 0cm}]{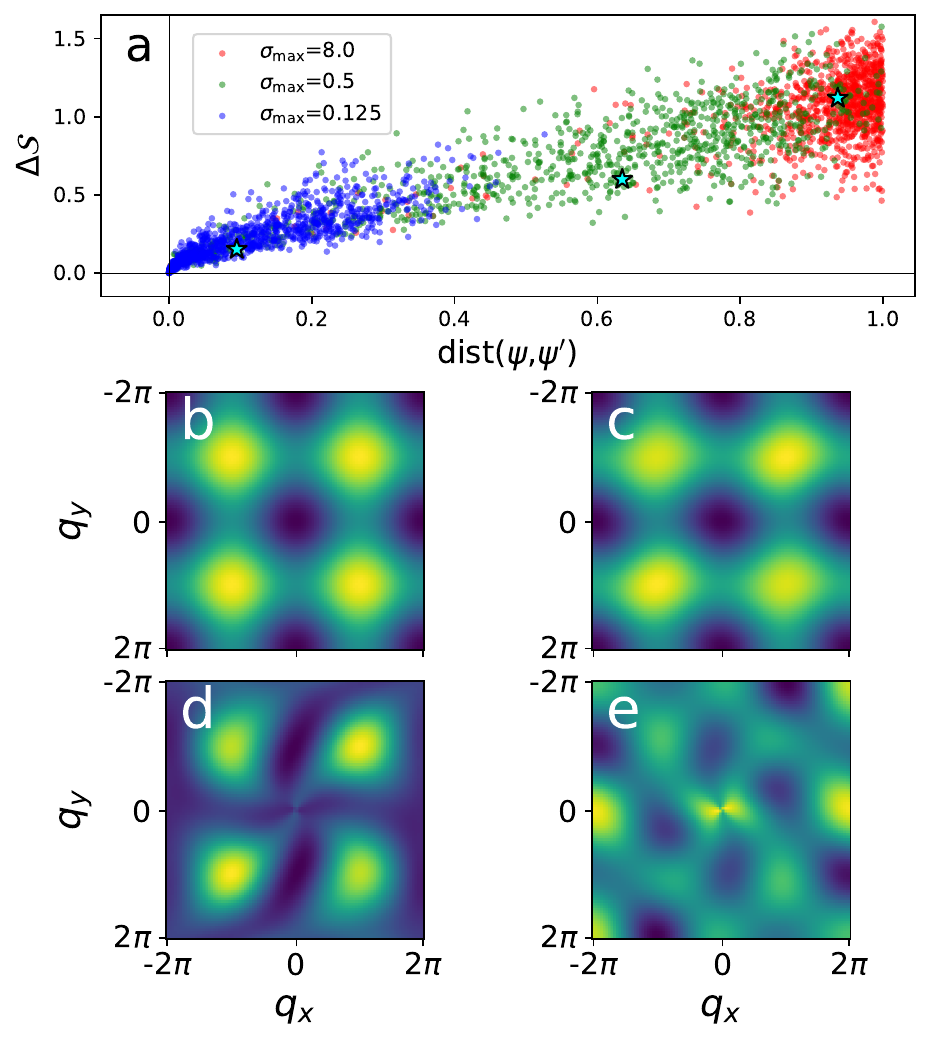} 
    \caption{Fitness landscape for quantum state tomography using the ground-state's diffuse magnetic neutron scattering function. The target state $\psi$ is the ground state of a spin-1/2 ring-tetramer (L=4) with nearest-neighbour, antiferromagnetic Heisenberg interactions. (a) Cost function $\Delta\mathcal{S}$ as a function of the distance ${\rm dist}(\psi,\psi')$ of the trial state vectors $\psi'$ to the target ground state \cite{supp-mat}. We display three sets of 1000 random state vectors each, with amplitudes drawn from Gaussian distributions centered on  $\psi$. The color legend indicates the maximum standard deviation $\sigma_\text{max}$ used for the set, while each trial state is marked by a filled circle and has components generated from distribution with a random width $\sigma$, where $0<\sigma<\sigma_\text{max}$ \cite{supp-mat}. (b) The target diffuse magnetic scattering function $\mathcal{S}\left(\mathbf{q}\right)$ for the exact ground state. (c-e) Three scattering functions obtained from the three random states highlighted with stars on panel (a), in order of increasing state-distance.}
    \label{fig:fitness_landscape_gs}
\end{figure}
We now present the results of our investigation into the fitness landscape for quantum state tomography in magnets. A first overview is given in Fig.~\ref{fig:fitness_landscape_gs}. In panel (a), we illustrate the relation between the distance $\Delta S$ of random states from the target scattering function against their state distance \cite{supp-mat} for a molecular magnetic tetramer ($L=4$) comprising four spin-$\frac{1}{2}$ described by the anti-ferromagnetic Heisenberg with nearest-neighbour interactions of fixed strength $J$ in a ring geometry. Random state vectors were created by adding Gaussian noise terms of width $\sigma$ to each component of the target state vector, followed by normalization \cite{supp-mat}. We display groups of results for three distinct maximal widths $\sigma_\text{max}$. Similar results were obtained using uniform noise distributions (not shown). Panel (a) shows a clear correlation between $\Delta\mathcal{S}$ and ${\rm dist}(\psi,\psi')$.

To gain some intuition, we also provide examples of scattering functions, including panel (b) for the target ground state and panels (c-e) for three of the randomly drawn states, hand-picked for representativeness [marked by stars in panel (a)]. The differences in the scattering functions are clearly appreciable by eye, even for the state (c) that is closer to the target. The scattering function in panel (c) is about 15\% from the target while the wave function overlap deviates by $\sim 5$\%. This suggests that it is possible to determine the system's ground state with considerable precision using realistically noisy neutron-scattering data. We also note that the candidate function in (c) has, to a very good approximation, the same four-fold symmetry as the target function, in contrast to the candidates with smaller overlaps, which look qualitatively different. This is a trend we see in all our simulations. These observations have practical implications: On the one hand, when symmetry constraints or other qualitative features of the data are known, including them can accelerate the search for the target state and improve its accuracy. On the other hand, our results show that an unbiased search through the Hilbert space may yield states with the correct features even if these are not known {\it a priori} and the experimental and target scattering functions can only be determined with limited accuracy.

\begin{figure}
    \centering
    \includegraphics[width=0.9\columnwidth,trim={0cm 0cm 0cm 0cm}]{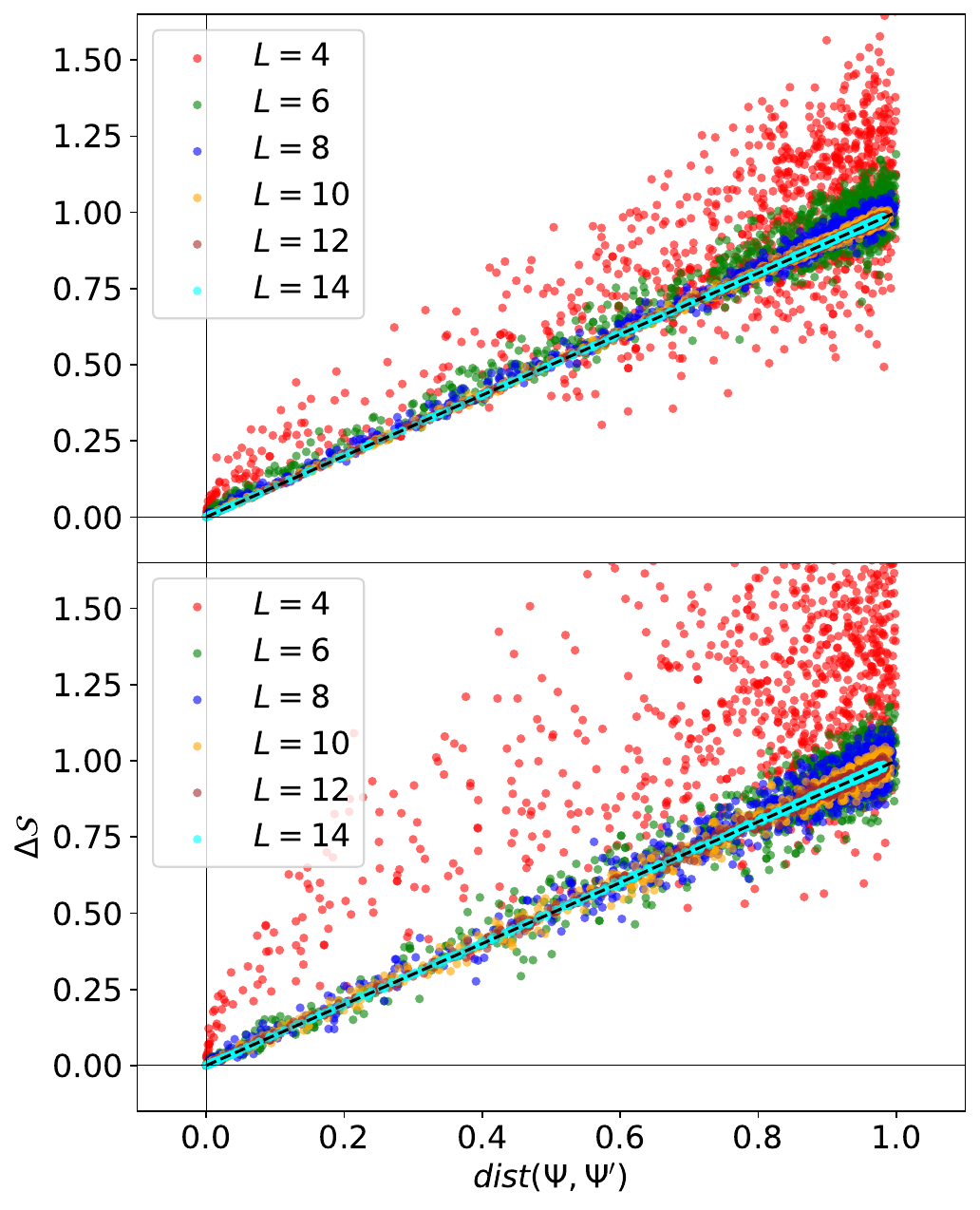}
	\caption{Dependence of the fitness landscapes on the number of spins $L$ in the cluster. For each $L$, 1000 target states were sampled with normal-distributed perturbations around the target wave function (see main text). Target states were chosen as ground states of: (a) A spin-1/2, nearest-neighbour anti-ferromagnetic Heisenberg model in a ring geometry, i.e., the same model as in Fig.~\ref{fig:fitness_landscape_gs}, but with varying L. (b) A random long-range spin-1/2 Hamiltonian in a linear open chain geometry, given by Eq.~(\ref{eq:HeisenbergHamField}) with random exchange constants \( J_{i,j}^{\alpha,\beta} \) and a random global magnetic field with components \( h_i^{\alpha} = h^{\alpha} \). For each $L$, Hamiltonian parameters \( J_{i,j}^{\alpha,\beta} \) and \( h^{\alpha} \) were drawn from a normal distribution with zero average and standard deviation equal to $J$ [the nearest-neighbour coupling of the Heisenberg Hamiltonian used in panel (a)]. Dashed lines in (a,b) show the linear relation in Eq.~(\ref{eq:straight_line}).}
	\label{fig:Detailed_fitness_gs-both}
\end{figure}

Next, we study antiferromagnetic Heisenberg rings of different lengths $L=4,6,8,10,12,14$, employing a size-dependent width in the noise distributions for good data coverage \cite{supp-mat}. Fig.~\ref{fig:Detailed_fitness_gs-both}(a) shows that as $L$ increases, the scatter in the data reduces, and they cluster around a single sharply defined curve. Remarkably, a simple linear relationship with coefficient equal to unity describes this curve very well:
\begin{equation}
	\Delta\mathcal{S}= {\rm dist}(\psi,\psi').
	\label{eq:straight_line}
\end{equation}  
Our results further suggest that trial states $\psi'$ approaching the target $\psi$ are more likely to obey the remarkably simple law (\ref{eq:straight_line}). This indicates an easy-to-navigate fitness landscape: once we are in the vicinity of the target scattering function, there is a well-defined slope towards to the target state. This is a remarkable result and constitutes our main finding. The fact that the ``empirical'' law (\ref{eq:straight_line}) is independent of $L$ encourages us to conjecture the relationship will hold even in the thermodynamic limit.

In the above discussion, we used the antiferromagnetic Heisenberg model with nearest-neighbour interactions to generate our target ground state. This was selected as a well-known textbook example for quantum magnets. The ring geometry was inspired by molecular quantum magnets which are experimental realisations of finite-size magnetic clusters amenable to treatment by exact diagonalisation and quantum eigensolvers~\cite{Schnack2023, kandala2017hardware}. However, our conclusions are robust to variations in the the model and geometry. To illustrate this, Fig.~\ref{fig:Detailed_fitness_gs-both}b) shows results for a linear molecule with randomly-chosen coupling constants and an externally-applied field of arbitrary strength and direction. In addition to the different geometry, these models lack SU(2) and time-reversal symmetries. Moreover, the couplings are defined on a fully-connected graph with no natural length scale. Thus, as $L$ grows the model can no longer be considered one-dimensional. Finally, we note that such random Hamiltonians typically have competing ferromagnetic and anti-ferromagnetic couplings. In spite of all these differences, we obtain very similar behavior. These results also confirm our proof shown above and in \cite{supp-mat} that the arguments in Ref.~\cite{Quintanilla2022} can be extended to pairwise-interacting spin models with an externally applied magnetic field. Moreover, they strongly suggest that the linear relationship (\ref{eq:straight_line}) describing the fitness landscape for quantum tomography in the vicinity of a target ground state applies universally. Our results in Fig.~\ref{fig:Detailed_fitness_gs-both}b) additionally demonstrate that the inverse scattering problem can be solved even for a non-periodic magnetic structure. We extend the arguments of Ref.~\cite{Quintanilla2022} and formally prove the possibility of quantum tomography for non-periodic structures or finite size clusters in the supplement \cite{supp-mat}.

{\it Finite temperature:} 
In the foregoing we have shown numerical evidence that that the $T=0$ ground states of pairwise coupled spins models can be inferred from neutron scattering data. Is the same true for magnetic systems at finite temperatures? In fact, while Hamiltonian learning from structure factors was proven to be possible at $T > 0$~\cite{Murta2022}, so far no extension for state tomography to thermal density matrices was found~\footnote{An analysis of the difficulty of proving this rigorously has been offered in Ref.~\cite{Murta2022}.}. Nevertheless, we can apply the same \emph{numerical} analysis as for the ground state, to check whether there is any numerical evidence to suggest that the relationship from finite-$T$ density matrices to the structure factor is also invertible. We test this hypothesis by sampling random, positive-definite density matrices in the vicinity of the thermal equilibrium density matrix of a given target Hamilonian with normal-distributed long-range couplings (details, see \cite{supp-mat}).

\begin{figure}
    \centering
    \includegraphics[width=0.9\columnwidth,trim={0cm 0cm 0cm 0cm}]{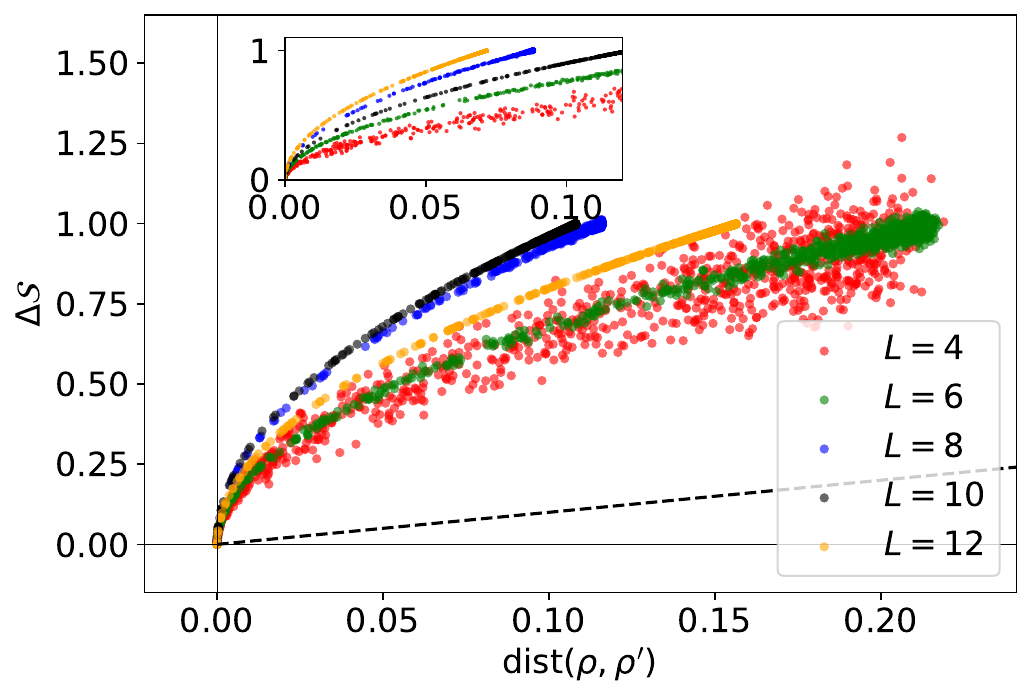} \\
    \caption{Dependence on $L$ (see legend) of the fitness landscape for finite-temperature quantum tomography for a model with random long-range interactions. The target Hamiltonian and geometry are defined in the same way as in Fig.~\ref{fig:Detailed_fitness_gs-both}b). The target density matrix $\rho$ is the thermal equilibrium density matrix of that Hamiltonian at temperature $T=J$. 1000 trial configurations (random density matrices) were tested against this target, based on adding noise terms followed by restoring positive definiteness and normalization \cite{supp-mat}. The horizontal axis shows a distance between random density matrices that for $T=0$ would have coincided with the overlap-induced distance $dist(\psi,\psi')$ between ground states \cite{supp-mat}. The dashed line represents  Eq.~(\ref{eq:straight_line}), for comparison. The inset shows similar output with different random targets at each size.}
    \label{fig:Finite_T_tomography}
\end{figure}

The resulting data is shown in Fig.~\ref{fig:Finite_T_tomography}, numerically supporting the hypothesis. The most salient feature, compared to the zero $T$ cases in Fig.~\ref{fig:Detailed_fitness_gs-both} is the much steeper (sub-linear) increase in $\Delta\mathcal{S}$ with increasing inter-state distance $\text{dist}(\rho,\rho')$. This is due to the fact that the space of trial  density matrices $\rho'$ is of higher dimensionality than that of trial pure states $\psi'$. Even if we choose the target state to be arbitrarily close to the ground state (by choosing the target temperature to be very low, compared to all the Hamiltonian parameters) we still find this markedly sub-linear behaviour, as the $\rho'$ matrices almost never represent pure states like those considered earlier.  Additionally, we find that the prefactor of the sub-linear scaling depends on the random choice for the target Hamiltonian (see the inset illustrating the data for a distinct set of randomized target Hamiltonians).

{\it Discussion and Conclusions:} In this paper we have presented an exact diagonalisation study of the fitness landscape for static magnetic structure factors of many-spin Hamiltonians with pairwise interactions, both in the ground state and at finite temperature. Our results allow us to assess the practicality of quantum state tomography based on diffuse magnetic neutron scattering. 

Our results include the important cases when there is a magnetic field applied to the sample. This was achieved through an analytical transformation between Hamiltonians with linear terms and purely quadratic Hamiltonians with complex interaction constants \cite{supp-mat}. This result extends the validity of our study and the theorems that underpin it~\cite{Quintanilla2022,Murta2022} and is our first result. 

Our results for the fitness landscape indicate that the cost functions for quantum tomography from structure factors are well behaved, both in the ground state and at finite temperature, with no evidence of barren plateaus or local minima. This is our second main result, and the core result in this paper. Thirdly, we proved that tomography is possible even from the structure factors of non-periodic magnetic clusters \cite{supp-mat}.

The case of quantum tomography at finite temperature deserves special attention. The proofs of the one-to-one correspondences between Hamiltonians and correlators (Theorem 1 in \cite{Quintanilla2022}) and between wave functions and correlators (Theorem 2) are based on the Rayleigh-Ritz variational principle \cite{becca2017quantum}, which is used by many numerical techniques, such as Density Functional Theory (DFT) \cite{capelle2006bird}, Variational Monte Carlo methods \cite{scherer2010computational} or recent Artificial Intelligence methods \cite{carleo2017solving}. The principle states that given a Hamiltonian $H$, the ground state will minimise the energy globally, which means that $E({\psi}) = \langle {\psi} | H | {\psi}\rangle / \langle {\psi} | {\psi} \rangle$ is a good optimization function for finding ground states. Their generalisations to systems at finite temperatures require using a similar inequality applied to thermal density matrices known as the Gibbs-Delbr\"uck-Moli\'ere quantum variational principle \cite{liu2021solving} (or, equivalently, the Gibbs-Bogoliubov inequality). In this case, the objective function is the free energy, defined by
\begin{equation} \label{eq:FreeEnergy}
    F({\rho}) = \mbox{Tr} \left[{\rho} H\right] - \beta^{-1} \mbox{Tr} \left[{\rho} \ln {\rho}\right],
\end{equation}
consisting of terms corresponding to expectation values of the energy and entropy respectively. The free energy (\ref{eq:FreeEnergy}) is minimised for the thermal density matrix ${\rho} = e^{-\beta H}/\mbox{Tr}\left[e^{-\beta H}\right]$. It turns out that the  entropy term does not alter the reasoning of the proof of Theorem~1, as shown  by Murta and Fern{\'a}ndez-Rossier~\cite{Murta2022}. However, as noted in that reference, that does not carry over to Theorem 2: one cannot discard the possibility that there are density matrices with lower entropy than the real one which reproduce the experimental data. However,  the systematic numerical exploration presented in the present paper did not provide any examples of this, giving strong indication that such cases, if they exist, must be special, at least for finite-size systems. This is the fourth principal result of the present study.

Finally, we note the advantage of using experimental neutron-scattering data as an input, rather than real-space correlators. At a mathematical level, both inputs are closely related due to their linear relation and arguments related to Parseval’s theorem of Fourier analysis (see App. A in \cite{Quintanilla2022} for the proof in the case of translational-invariant systems and \cite{supp-mat} for the extension to finite clusters), so our approach would operate similarly on both input data. However, experimental approaches giving access to the spatial two-site correlation functions, including experiments with cold atoms / ions \cite{ottSingleAtomDetection2016, parsonsSiteresolvedMeasurementSpincorrelation2016, ColdAtomsCorrelations} or potentially surface measurements using local probes \cite{willkeHyperfineInteractionIndividual2018, elbertseRemoteDetectionRecording2020, DblResonance, CertEntESR-STM}, typically require separate measurements for each pairwise correlator and, for the latter, probe only surface states. In contrast, as discussed in \cite{Quintanilla2022}, neutron scattering yields information about the bulk wave function of the material in a single shot.

Our results suggest that the variational approach to quantum magnets proposed in Ref.~\cite{Quintanilla2022}, where an efficient representation of the wave function of a magnet is optimized to reproduce experimentally-obtained neutron scattering data, may be feasible, at least for magnets formed of finite-size clusters. Additional studies will be necessary to ascertain whether this conclusion carries over to fully-connected systems as well as the effect of noisy and incomplete data sets. Effective ways to represent the wave function, for instance using neural networks, tensor networks or quantum circuits (and the potential of the latter to achieve scaling advantages when implemented on noisy, intermediate-scale quantum hardware) will also need to be developed. 

\begin{acknowledgments}
GM acknowledges insightful discussions with F. Verstraete, who pointed out that our results are supported by a geometric interpretation, namely that the ensemble of possible expectation values in thermal equilibrium form a convex set in observable space \cite{Zauner2016}.
TT is supported by the EPSRC via a DTA studentship under Grant No. EP/R513246/1 and by the School of Physical Sciences, University of Kent. GM acknowledges support by the Royal Society under University Research Fellowship award URF\textbackslash R\textbackslash 180004.

{\it Code availability: } The computer codes used to generate the data supporting the findings in this article are openly available \cite{FitnessLandscapeCode}.

\end{acknowledgments}

\bibliographystyle{apsrev4-2}
\bibliography{bibliography}

\appendix
\input{SuppMat_sections}

\end{document}

%% file: SuppMat_sections.tex
\section{\label{sec:approach_detail}Approach to sampling the fitness landscape: Detailed description}

Here we provide full detail on the procedure used to test the fitness landscape of scattering functions in a space of wave functions for small model Hamiltonians. To generate the data presented in our study, we proceed in the following manner:\newline
(1) A reference Hamiltonian $H$ is selected which describes a chain of $L$ sites of $S=\frac{1}{2}$ spins. We consider two types of geometries: Either we will assume the geometry of the simple model considered in Ref.~\cite{Irons2017}: $L$ sites forming a planar ring molecule with $C_L$ rotational symmetry (and hence Hamiltonian parameters invariant under translations of the site indices $i,j$). Alternatively, we also considered, where noted, the case of a linear molecule with open boundary conditions. 
For our $T=0$ studies, we obtain the target ground state $\psi$ through exact diagonalisation. Alternatively, at finite temperature $T$, we compute the thermal equilibrium density matrix, $\rho = \mbox{Tr}{\left[e^{-\beta H}\right]}^{-1}e^{-\beta H}$, from the full spectrum. Once the state is known, the two-point spin correlation functions $\langle S_i^\alpha S_j^\beta\rangle$ are evaluated, 
and the corresponding instantaneous magnetic structure factors $\mathcal{S}_{\alpha,\beta}(\mathbf{q})$ are calculated as follows: 
    For scattering vectors $\mathbf{q}$ contained in the plane of the molecule, the structure factor is given by 
        \begin{multline}
            \label{eq:nsfunctions}
            \mathcal{S}_{\alpha,\beta}(\mathbf{q})
            =
            \frac{1}{L\hbar}\sum_{i,j}
            e^{i\mathbf{q}\cdot
                \left(
                    \mathbf{R}_i-\mathbf{R}_j
                \right)
            } \times  
            \\
            \times  \sum_{\mu,\nu}
            \Lambda_{\alpha,\mu}\left(\phi_i\right)
            \Lambda_{\beta,\gamma}\left(\phi_j\right)
            \langle S_i^\mu S_j^\gamma\rangle.
        \end{multline}
Here $j\in{1,\ldots,L}$ is the site index,  $\mathbf{R}_j=R\,{\left(\sin\phi_j,\cos\phi_j\right)}^t$ is the position vector of the $j^{\underline{\text{th}}}$ magnetic site within the molecule where $R$ is the radius of the molecule and angles are equally spaced as  $\phi_j=\frac{\pi}{L}\left(1-2j\right)$.  The matrix ${\left\lbrace\Lambda_{\alpha,\mu}\left(\phi_j \right)\right\rbrace}_{\alpha=x,y,z \atop \mu=x,y,z}$ rotates the $j^{\underline{\text{th}}}$  spin from the laboratory frame to its local axes.  
        We note that the distance between nearest-neighbour magnetic sites is $a=2R\sin(\pi/L)$. The rotation matrices and other details of the model are given explicitly in Ref.~\onlinecite{Irons2017}. For the case of a linear molecule, there is no need to rotate the axes.
{The full diffuse magnetic neutron scattering function is then obtained \textit{via} 
\begin{equation}
    \mathcal{S}(\mathbf{q})
    =
    \sum_{\alpha.\beta}
    \left(
        \delta_{\alpha,\beta}-\hat{\mathbf{q}}_{\alpha}\hat{\mathbf{q}}_{\beta}
    \right)
    \mathcal{S}_{\alpha,\beta}(\mathbf{q}).
\end{equation}
(2) We generate a set number of random states $\psi'$ that are different from the ground state $\psi$ of $H$ (or, at finite temperature, random density matrices $\rho'$). This can be achieved either by sampling random Hamiltonians $H'$ and calculating their ground states (or thermal density matrices), which restricts the exploration of the Hilbert space, or through adding a random variation to the ground state, $\psi' = \psi + \delta \psi$ (or to the density matrix, $\rho' = \rho + \delta \rho$), to be properly normalised. Our procedures for random samplings in these cases are further explained in supplementary note \ref{sec:detail_random_sampling}.\\
(3) We calculate a metric distance in the space of structure factors $\Delta \mathcal{S}$ between the reference state and each of the randomly generated scattering functions $\mathcal{S},\mathcal{S}'$ as defined in supplementary section \ref{sec:metrics_detail_SF}. We then examine how it relates to the distance between the ground state $\psi$ of our reference Hamiltonian and the random state  $\psi'$ (or between the respective density matrices). The appropriate metrics to evaluate distances in state-space are presented in supplementary section \ref{sec:metrics_detail_states}.

\section{\label{sec:metrics_detail_SF}Choice of metric for measuring distances between structure factors}

To quantify the distance between scattering functions we use the root mean square (r.m.s.)~deviation 
\begin{equation} \label{eq:functionalS}
    \Delta \mathcal{S}=\Delta \mathcal{S}[\mathcal{S}, \mathcal{S}'] = 
    \frac{1}{\mathcal{N}}
    \left\lbrace 
        \int d\mathbf{q} {\left[\mathcal{S}(\mathbf{q}) - \mathcal{S}'(\mathbf{q})\right]}^2
    \right\rbrace^{1/2}.
\end{equation} 
In order to make the absolute values of $ \Delta \mathcal{S} $ meaningful, a judicious choice of the normalisation factor $\mathcal{N}$ is necessary. We take 
\begin{equation} \label{eq:functionalSN}
    \mathcal{N} = 
    {\left\lbrace
        \int d\mathbf{q} {\left[\mathcal{S}(\mathbf{q}) - \bar{\mathcal{S}}\right]}^2
    \right\rbrace}^{1/2}
\end{equation} 
where $\bar{\mathcal{S}}$ is the average of $\mathcal{S}\left(\mathbf{q}\right)$ over all possible momentum transfers $\mathbf{q}$.

We note that our chosen normalization $\mathcal{N}$ is defined using the target scattering function $S(\mathbf{q})$, which appears to be an asymmetric choice. However, if the domain of the momentum integration covers all of $\mathbf{q}$-space, the exact sum rule $\bar{\mathcal{S}}=s(s+1)$ holds~\cite{Broholm} and our expressions simplify.
The sum rule yields the same numerical value irrespective of whether $S(\mathbf{q})$ or $S'(\mathbf{q})$ is used to evaluate the right hand side of Eq.~(\ref{eq:functionalSN}) ensuring that the value of $\Delta \mathcal{S}$ is independent of the order in which we write the scattering functions. We thus obtain a metric in the strict mathematical sense, satisfying $\Delta \mathcal{S}[\mathcal{S}, \mathcal{S}'] = \Delta \mathcal{S}[\mathcal{S}', \mathcal{S}]$.
However, in practice we often have experimental information available only for a restricted domain in momentum space, caused for instance by the detector coverage of the neutron-scattering instrument. In this case we choose the normalisation against the reference state, and the resulting distance-measure is not strictly a metric in the mathematical sense.
%

With our chosen normalisation \eqref{eq:functionalSN}, a value of  $\Delta \mathcal{S} \sim 1$ indicates that the deviation of the trial scattering function $\mathcal{S}'(\mathbf{q})$ from the target one $\mathcal{S}(\mathbf{q})$ is appreciable on the scale of the features of the target function itself. As it turns out with this choice of normalisation we obtain a remarkable independence of our loss function from the system size $L$ in the $T=0$ case (see main text).

\section{\label{sec:metrics_detail_states}Choice of metrics in state-space}
In the state tomography problem, the distance $\Delta\mathcal{S}$ between scattering functions needs to be compared to a distance between the ground state $\psi$ and the target state $\psi'$ (or between density matrices $\rho$ and $\rho'$). For pure states, it is straightforward to define a metric for states via the overlap as
\begin{equation} \label{eq:dist_psi}
    \text{dist}(\psi, \psi') = 1 - \left|\langle \psi | \psi' \rangle\right|^2.
\end{equation}
We take its natural generalisation for mixed states as the trace of their square distances,
\begin{equation} \label{eq:dist_rho}
    \text{dist}(\rho, \rho') = \frac{1}{2} \text{Tr} \left[{\left(\rho-\rho'\right)}^\dagger \left(\rho - \rho'\right)\right],
\end{equation}
which reduces to (\ref{eq:dist_psi}) when $\rho=|\psi\rangle\langle\psi|$ and $\rho'=|\psi'\rangle\langle\psi'|$.}

\section{\label{sec:detail_random_sampling}Generation of Random States}

\subsection{\label{sec:detail_random_sampling_states}Sampling eigenstates ($T=0$)}
We consider states of the form $|\Psi'\rangle = |\Psi	 + R\rangle$, where random entries drawn from a normal distribution with some standard deviation $\sigma$ are added to each component of the target state, and the resulting vector is then normalised. Written more explicitly in component-wise notation, this corresponds to the following construction:
\begin{enumerate}
\item Draw real parts and imaginary parts for the components of $R_j$ according to the distribution $p[\Re(R_j)] = p[\Im(R_j)] = N(\sigma,0)[x]$, for $j=1,\ldots,\dim(\mathcal{H})$, with $N(\sigma,\mu)$ the normal distribution of standard deviation $\sigma$ and mean $\mu$, and we denote the Hilbert space dimension of the problem as $\dim(\mathcal{H})$.
\item Construct the unnormalised perturbed state with components $\tilde \Psi'_j = \Psi_j + R_j$.
\item Normalize the resulting wave function $\Psi'_j = \mathcal{N} \tilde \Psi'_j$, choosing $\mathcal{N}$ to enforce $\sum_j \Psi'_j=1$, yielding the components of the new ket $|\Psi'\rangle$.
\end{enumerate}

Since the sum of squares of $n=2\dim(\mathcal{H})$ Gaussian distributions yields a $\chi^2(n)$ distribution with expectation value $E(\chi^2(n)) = n$, our sampling procedure yields noise vectors $\vec R$ whose typical norm of order  $E(|R|) = \sigma \sqrt{n}$. Hence, the average deviation from the starting state has magnitude proportional to $\sqrt{\dim(\mathcal{H})}$. Although normalization removes one degree of freedom, we expect that the typical distance for states sampled with a fixed $\sigma$ yields an overlap of the order of $1-\sigma \sqrt{2\dim(\mathcal{H})}$ from the target if the perturbation is sufficiently small (and saturating to $1$ if it is large). This observation has two important consequences.

Firstly, a given choice of $\sigma$ tends to result in perturbed states $\psi'$ with a similar distance from the starting state, since $\text{var}(\chi^2(n))=2E(\chi^2(2n))$ and the magnitude of fluctuations around the average norm of the random term shrinks with system size. To ensure even coverage of random states both close to and farther away from our target state, we therefore choose a distinct standard deviation $\sigma$ for each trial state which we choose from a uniform distribution $\sigma\in [0,\sigma_\text{max})$. Then we proceed to generate all $n$ components of that state with the Gaussian distribution $N(\sigma,0)$.

Secondly, when comparing results for several system sizes, we see that the typical distance is dominated by the number of random choices made in $R$, i.e.~$2\dim\mathcal{H}$. Hence, we additionally choose to scale the standard deviation $\sigma$ used for individual samples in a system-size dependent way, taking the range of its distribution as $\sigma_{\max}(L)=\sigma_{\max} \times (\dim\mathcal{H(L)})^{-\frac{1}{2}} = \sigma_{\max} \times 2^{-\frac{L}{2}}$, to ensure that the sampled random vectors cover roughly the same range of distances for all system sizes.

{\it Other distributions:} The Gaussian distribution can be replaced by a uniform distribution. In the latter case, drawing the real and imaginary parts independently yields distribution within a square in the complex plane, so we prefer a radially symmetric choice uniformly sampling complex numbers $R_j$ within a disc of radius $\sigma$. We have tested that both flavours of uniform distributions yield qualitatively similar results to the choice of normal distributed noise.

\subsection{\label{sec:detail_random_sampling_density_matrices}Sampling density matrices ($T\neq 0$)}

 To randomly sample trial density matrices $\rho'$ in the vicinity of the target $\rho$, we have explored two distinct strategies. In either case, we start from a random symmetric matrix $R$ with random Gaussian-distributed entries of width $\sigma$, and proceed in one of the following two ways, by either
 \begin{itemize}
 \item ensuring positive definite eigenvalues by squaring the noise term, corresponding to the choice $$\rho' = \mathcal{N}(\rho + R^\dagger R),$$ with the normalization constant $\mathcal{N}$ chosen such that $\tr(\rho')=1$, or alternatively 
 \item eliminating negative eigenvalues from the spectral decomposition of $\tilde M = \rho + R = U D U^\dagger$, where $U$ is the matrix of eigenvalues, and $D = \text{diag}(D_1,\ldots, D_{\dim(\mathcal{H})})$ is the diagonal matrix of the corresponding eigenvalues $D_i$. Then a positive definite matrix is obtained by replacing $D\to \tilde D$, with $$\tilde D= \text{diag}(\max(D_1,0),\ldots, \max(D_{\dim(\mathcal{H})},0)).$$ Again, we include a normalization factor and define $\rho' = \mathcal{N}' U \tilde D U^\dagger$, with $\mathcal{N}'$ such that $\tr(\rho')=1$. 
 \end{itemize}
 In practice, we find that both approaches give similar results. However, the first approach is more numerically scalable, so it was chosen to generate the data presented in the main text. 

 The randomness generated in this procedure has of order $\dim(\mathcal{H})^2$ degrees of freedom, since $\rho$ is a $\dim(\mathcal{H}) \times \dim(\mathcal{H})$ matrix. Thus, we require system-size scaling $\propto (\dim\mathcal{H}(L))^{-\frac{1}{4}}$. In practice we find that this factor does not capture all of the size-variation, so we additionally insert an empirically chosen size-dependent factor to obtain similar sampling ranges of the state distance across all system sizes. Specifically, we take $\sigma \in [0, c_L \sigma_{\max} \times 2^{-\frac{L}{4}})$, where we have used $c_4=c_6=c_8=0.3$, $c_{10}=0.1$, and $c_{12}=0.03$ for Fig.~\ref{fig:Finite_T_tomography} in the main text.
 
 Again, we can replace the Gaussian distribution for noise elements with other distributions, and as for the zero-temperature case, we continue to observe qualitatively the same results when doing so.

\section{\label{sec:linear_terms}Generalisation to systems with linear spin terms}
The general form of the Hamiltonian $H_{J}$  in Eq.~(\ref{eq:HeisenbergHam})\footnote{Equations labeled by roman numbers refer to the main text.} encompasses many different magnetic systems including the Ising model \cite{ising1924beitrag}, XXZ model \cite{gu2005ground}, Kitaev model \cite{kitaev2006anyons}), and many models with long-range interactions. However, the lack of linear spin terms excludes the effects of external (or local) magnetic fields. A more general model would take the form given in Eq.~(\ref{eq:HeisenbergHamField}). In the following, we demonstrate that the two classes are equivalent, provided that we allow the $J_{i,j}^{\alpha, \beta}$ coefficients in (\ref{eq:HeisenbergHam}) to be complex. We will do so by deriving an explicit transformation between $H_{J}$ and $H_{J+h}$.  

Our starting point are the commutation relations for the three components of the spin at a given site $j$: 
\( \left[ S_j^{\alpha}, S_j^{\beta}\right] = i\hbar \sum_{\gamma=x,y,z} \epsilon_{\alpha\beta\gamma}S_j^{\gamma}, \) where $\epsilon_{\alpha\beta\gamma}$ is the Levi-Civita symbol. Taking $\alpha\beta\gamma=yzx, zxy$ and $xyz$, the right-hand side takes the form of the constant $i\hbar$ multiplied by $S_j^x, S_j^y$, and $S_j^z$, respectively. We can use these relations to recast the spin operators in the second term on the right-hand side of Eq.~(\ref{eq:HeisenbergHamField}) as sums of bilinears, which can be absorbed into the pre-existing bilinear terms. The result reads 
\begin{equation} \label{eq:2primed}
    H_{J+h} = \sum_{i,j}\sum_{\alpha, \beta} \bar{J}_{i,j}^{\alpha, \beta} S_i^{\alpha} S_j^{\beta},
\end{equation}
where we have introduced the new, modified exchange constants
\begin{equation}
\label{eq:replacement_of_Js}
    \bar{J}_{i,j}^{\alpha,\beta} 
    \equiv 
    {J}_{i,j}^{\alpha, \beta} - \frac{i}{2\hbar}\delta_{i,j}\epsilon_{\alpha\beta\gamma}h_j^{\gamma}.
\end{equation}
Here $\alpha,\beta=x,y,z$ and $\alpha\neq\beta$, while $\gamma$ takes the value that makes $\alpha\beta\gamma$ a permutation of $xyz$, so two of these terms contribute for each choice of $\gamma$. 

Note that with the above definition, the effective exchange constants $\bar{J}_{i,j}^{\alpha,\beta}$ can be complex. We still have $\bar{J}_{i,j}^{\alpha,\beta}=\left(\bar{J}_{j,i}^{\beta,\alpha}\right)^*$, which follows from ${J}_{i,j}^{\alpha,\beta}={J}_{j,i}^{\beta,\alpha}$ and $h_j^{\alpha}$ being real and keeps $H_{J+h}$ being Hermitian. Let us review how this modification affects the proofs by {\it reductio ad absurdum} of Hamiltonian learning and quantum state tomography presented in~\cite{Quintanilla2022}.  

The premise of both proofs is that we have equal correlators for two different states at the outset, that is, $\correlator_{ij}^{\alpha\beta}
[\Psi] = 
\correlator_{ij}^{\alpha\beta}
[\Psi']$, for all $i,j,\alpha,\beta$. 
In proving the 
possibility of Hamiltonian learning
, the two states $\Psi_0$ and $\Psi_0'$ are assumed to be ground states of two different Hamiltonians. We arrive at two inequalities of the form 
\begin{align}
    E_0 + E'_0  < & \sum_{i,j,\alpha,\beta} ({J'}_{ij}^{\alpha\beta}-J_{ij}^{\alpha\beta})\left[\correlator_{ij}^{\alpha\beta}[\Psi_0] - \correlator_{ij}^{\alpha\beta}[\Psi_0']\right] \nonumber \\ 
    & + E_0 + E_0',
\end{align}
and an equivalent relation with the 
smaller-than sign replaced by a 
larger-than sign. In these expressions, the couplings will be replaced according to \eqref{eq:replacement_of_Js}, introducing some complex terms. However, according to the initial assumption, the term in the square brackets vanishes, so the additional imaginary terms get multiplied by zero. Hence, the reasoning in the proof is not affected.

In the quantum state tomography proof, the two examined states
are the ground state $\Psi_0$ of the target Hamiltonian and an arbitrary state $\tilde \Psi$. We are are then led to consider the inequality 
\begin{align}
    E_0 = \langle\Psi_0|H|\Psi_0\rangle  & = \sum_{i,j,\alpha,\beta} J_{ij}^{\alpha\beta}\correlator_{ij}^{\alpha\beta}[\Psi_0]  \nonumber \\
    & \leq 
    \sum_{i,j,\alpha,\beta} J_{ij}^{\alpha\beta}\correlator_{ij}^{\alpha\beta}[\tilde \Psi].
\end{align}
We may worry that after replacing the $J$ coupling constants with the $\tilde{J}$ ones in Eq.~\eqref{eq:replacement_of_Js}, the above sums could become complex. However, by Hermitian symmetry of the Hamiltonian, the energy eigenvalues remain real. Furthermore, the assumption of equal correlators, which  in this case takes the form $\correlator_{ij}^{\alpha\beta}[\Psi_0] = \correlator_{ij}^{\alpha\beta}[\tilde\Psi]$ for all $i,j,\alpha,\beta$, explicitly includes the on-site terms with $i=j$. It is these terms which encode the magnetic fields, so the inequality leading to the {\it reductio ad absurdum} still holds and the proof proceeds as before.


\section{\label{sec:ProofForFiniteClusters}Proof of equivalence between knowledge of the 2-point correlators
and of the spin-resolved diffuse scattering function for finite clusters}

In this section we extend the reasoning of Ref.~\cite{Quintanilla2022}, which explicitly only proved the possibility of quantum state tomography and Hamiltonian learning from the magnetic structure of translational-invariant systems. We are not aware of a prior proof in the case of non-periodic systems, so we hereby provide a formal proof:

\noindent{\it Proof:} Consider two scattering functions 
\begin{subequations}
\begin{align}
S_{\alpha,\beta}^{1}\left(\mathbf{q}\right)=\sum_{i,j}e^{i\mathbf{q}\cdot\left(\mathbf{R}_{i}-\mathbf{R}_{j}\right)}\correlator{}_{i,j}^{\alpha,\beta}{\left[\Psi_{1}\right]}\label{eq:S1}\\
S_{\alpha,\beta}^{2}\left(\mathbf{q}\right)=
\sum_{i,j}e^{i\mathbf{q}\cdot\left(\mathbf{R}_{i}-\mathbf{R}_{j}\right)}
\correlator_{i,j}^{\alpha,\beta}{\left[\Psi_{2}\right]}\label{eq:S2}
\end{align}
\end{subequations}
corresponding to the correlation functions $\correlator_{i,j}^{\alpha,\beta}$ of two different quantum states $|\Psi_{1}\rangle,|\Psi_{2}\rangle$:
\begin{subequations}
\begin{align}
\correlator_{i,j}^{\alpha,\beta}\left[\Psi_{1}\right] & =\left\langle \Psi_{1}\left|S_{i}^{\alpha}S_{j}^{\beta}\right|\Psi_{1}\right\rangle \\
\correlator_{i,j}^{\alpha,\beta}\left[\Psi_{2}\right] & =\left\langle \Psi_{2}\left|S_{i}^{\alpha}S_{j}^{\beta}\right|\Psi_{2}\right\rangle
\end{align}
\end{subequations}
Let us further assume that we are dealing with a finite cluster with only
a finite number of sites whose position vectors 
\[
\mathbf{R}_{1},\mathbf{R}_{2},\ldots,\mathbf{R}_{N}
\]
are known. The wave vector $\mathbf{q}=\mathbf{k}_{f}-\mathbf{k}_{i}$,
on the other hand, is just the change in momentum of the neutron after
it interacts with the cluster and therefore takes continuous values.
We wish to prove that if $S_{\alpha,\beta}^{1}\left(\mathbf{q}\right)$
and $S_{\alpha,\beta}^{2}\left(\mathbf{q}\right)$ are identical, this implies that all the $N^{2}$ correlators coincide (that
is, $\correlator_{i,j}^{\alpha,\beta}\left[\Psi_{1}\right]=\correlator_{i,j}^{\alpha,\beta}\left[\Psi_{2}\right]$
for all $i,j\in\{1,2,\ldots,N\}$).

It is helpful to re-write (\ref{eq:S1}) and (\ref{eq:S2}) in the
following shorthand notation: 
\begin{align}
S^{1}\left(\mathbf{q}\right) & =\sum_{n}e^{i\mathbf{q}\cdot\mathbf{r}_{n}}\correlator_{n}^{1}\label{eq:S1-1}\\
S^{2}\left(\mathbf{q}\right) & =\sum_{n}e^{i\mathbf{q}\cdot\mathbf{r}_{n}}\correlator_{n}^{2}\label{eq:S2-1}
\end{align}
Here we have omitted the dependence on $\alpha,\beta$ and we have
introduced an abbreviated notation where $n$ is an index that runs
through all $N^{2}$ combinations of $i$ and $j$ [so, for example,
$n=1$ represents $\left(i,j\right)=\left(1,1\right)$, $n=2$ represents
$\left(1,2\right)$, and so on up to $n=N^{2}$ which represents $\left(N,N\right)$]
and the vector $\mathbf{r}_{n}\equiv\mathbf{R}_{i}-\mathbf{R}_{j}$
connects one site within the cluster to another one. 

Let us now regard $e^{i\mathbf{q}.\mathbf{r}_{n}}$ as a plane wave
with $\mathbf{q}$ as the independent (continuous) variable. Then,
with this language, $\mathbf{r}_{n}$ plays the role of the wave vector
which, mathematically speaking, could also take any value, although
in our problem it only takes specific values given by the geometry
of our magnetic cluster. It is clear nevertheless from Eqs.~(\ref{eq:S1-1},\ref{eq:S2-1})
that $S^{1}\left(\mathbf{q}\right)$ and $S^{2}\left(\mathbf{q}\right)$
are two linear combinations of the same $N^{2}$ $\mathbf{q}$-dependent
``plane waves'' with fixed ``wave vectors'' $\mathbf{r}_{1},\mathbf{r}_{2},\ldots,\mathbf{r}_{N^{2}}$.
Now, the set of all plane waves $\left\{ e^{i\mathbf{q}.\mathbf{r}}\right\} _{\mathbf{r}\in\mathbb{R}^{3}}$
form a complete, orthogonal basis that can describe any function
of $\mathbf{q}$. Specifically, we can represent the functions in terms of their Fourier modes:
\begin{align*}
S^{1}\left(\mathbf{q}\right) & =\int d^{D}\mathbf{r}\,e^{i\mathbf{q}\cdot\mathbf{r}}\correlator^{1}\left(\mathbf{r}\right)\\
S^{2}\left(\mathbf{q}\right) & =\int d^{D}\mathbf{r}\,e^{i\mathbf{q}\cdot\mathbf{r}}\correlator^{2}\left(\mathbf{r}\right)
\end{align*}
Evidently, if $\correlator^{1}\left(\mathbf{r}\right)$ and $\correlator^{2}\left(\mathbf{r}\right)$
are the same function then $S^{1}\left(\mathbf{q}\right)$ and $S^{2}\left(\mathbf{q}\right)$
are the same function too. Since these expressions are nothing but Fourier transforms, which are invertible, it follows by taking the inverse transform on both sides of these equations that the equality of the $S(\mathbf{q})$'s likewise implies the equalities of the $\correlator(\mathbf{r})$'s.
But Eqs.~(\ref{eq:S1-1},\ref{eq:S2-1}) can be brought into this form, by setting
\begin{align*}
\correlator^{1}\left(\mathbf{r}\right) & =\sum_{n}\delta\left(\mathbf{r}-\mathbf{r}_{n}\right)\correlator_{n}^{1}\\
\correlator^{2}\left(\mathbf{r}\right) & =\sum_{n}\delta\left(\mathbf{r}-\mathbf{r}_{n}\right)\correlator_{n}^{2}
\end{align*}
So evidently the two functions $S^{1}\left(\mathbf{q}\right)=S^{2}\left(\mathbf{q}\right)$
if, and only if, $\correlator_{n}^{1}=\correlator_{n}^{2}$, \emph{Q.E.D.}